# Electronic Laboratory Notebook on Web2py Framework

Yong-Yao Ng[1,2], Maurice HT Ling[2,3,4]
[1]School of Media and Information Technology, Singapore Polytechnic
[2]School of Chemical and Life Sciences, Singapore Polytechnic
[3]Department of Zoology, The University of Melbourne, Australia
[4]Corresponding author: mauriceling@acm.org

**Abstract**

Proper experimental record-keeping is an important cornerstone in research and development for the purpose of auditing. The gold standard of record-keeping is based on the judicious use of physical, permanent notebooks. However, advances in technology had resulted in large amounts of electronic records making it virtually impossible to maintain a full set of records in physical notebooks. Electronic laboratory notebook systems aim to meet the stringency for keeping records electronically. This manuscript describes CyNote which is an electronic laboratory notebook system that is compliant with 21 CFP Part 11 controls on electronic records, requirements set by USA Food and Drug Administration for electronic records. CyNote is implemented on web2py framework and is adhering to the architectural paradigm of model-view-controller (MVC), allowing for extension modules to be built for CyNote. CyNote is available at http://cynote.sf.net.

## 1. Introduction

Since the creation of the written languages, research work and the thoughts behind it are documented as written words and diagrams in physical forms which can be clay tablets in ancient times or bounded paper notebooks as of today. Proper record-keeping is paramount in the research and development arena as it is crucial to assert the date of any invention or discovery (Garabedian, 1997). At the same time, these records are to be presented as evidence of discovery or invention in events of dispute (Garabedian, 2003). The Director's office of United States of America (USA) National Institute of Health (NIH) had recently provided 5 reasons for proper record-keeping in scientific research, as follows (US NIH, 2008):

- Good record keeping is necessary for data analysis, publication, collaboration, peer review, and other research activities.
- Good record keeping is required by the NIH to meet the accepted policies and standards for the conduct of good science.
- Good record keeping is necessary to support intellectual property claims.
- Good record keeping can help defend you against false allegations of research misconduct.
- Good record keeping is important in the care of human subjects.

This implies that dating of entries is highly important and any actions that may lead to suspicion of record tampering; such as blank pages, pages dated out of order, removal of pages or corrections which destroy original entries (removal of pages or blotting out); are prohibited. Extensive works on proper record-keeping procedures (Kanare, 1985; Schreier et al., 2006) and record storage (William et al., 2008) had been published as a guide.

However, with technological advances, documentation using only bounded paper notebooks appears to be inadequate as a research project today can be littered with numerous digital documents in the forms of images and discussion emails to name a few. In order to formalize the requirements of electronic record-keeping, USA Food and Drug Administration (FDA) had gazetted 21 Code of Federal Regulations Part 11 (commonly known as 21 CFR 11), entitled "Electronic Records; Electronic Signatures" (US FDA, 1997), which specifies the control requirements for electronic records and electronic signatures.

This manuscript describes the development of CyNote - a 21 CFR 11 compliant electronic laboratory notebook to cater to the needs of student projects within Singapore Polytechnic, as well as for general research needs. The current version of CyNote is 1.4, released under GNU General Public Licence 3, and can be downloaded from http://cynote.sourceforge.net.

2.    **Architecture of CyNote**

CyNote is built on web2py (Di Pierro, 2009), a Python web framework, using default application structure and adhering to the model-view-controller (MVC) paradigm. Web2py is capable of running different applications within a single instance and maps a URL to its respective Python function. For example, the URL "http://localhost:8000/cynote/account/newaccount" is mapped to the *newaccount* function found within the *account* controller of *cynote* application (Figure 1). Each controller function will require its own rendering template, in the form of a HTML file, under the views/<controller> directory.

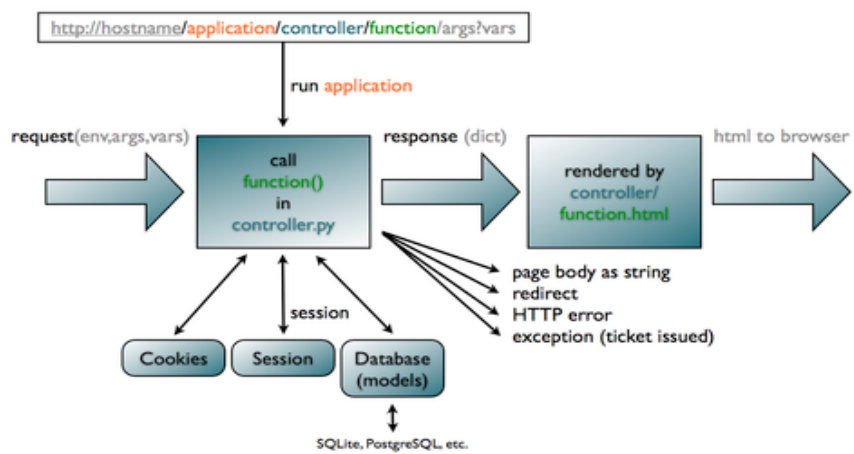

Figure 1: Web2py URL Mapping. Adapted from Di Pierro (2009).

CyNote stores all its data into one of the three SQLite (Newman, 2004) databases, namely; *cynotedb*, for storing notebook data and analysis data; *db*, for logging and auditing purposes and *userdb*, for storing user identity. A total of 7 controller modules were implemented in CyNote – *default*, *account*, *cynote*, *primer*, *savedatabase*, *sequence* and *statistics* – based on their functionalities. The modules, *default*, *account*, *cynote* and *savedatabase*, implement note-taking and administrative features while *primer, sequence* and *statistics* implement the add-on computing features.

**Module: *default*.** This module is the entry point into CyNote and shows the main menu screen. It performs three checks in the background. Firstly, it will check for the presence of BioPython (Cock et al., 2009) module and attempts to install it using *setuptools* if BioPython is not present. Secondly, it will check for user login status. If no user is logged in, control will be directed for user to log into the system. Lastly, it will check for the age of user password. If the password is more than 90 days old, control will be directed for the user to change his/her password.

**Module: *account*.** This module manages user account information. It contains functions for users to create new account, logging into and out of the system, changing password, authorizing and de-authorizing user accounts. Each of these events will be logged in *db* database.

**Module: *cynote*.** This module implements note-taking features which manages the cynote database which keeps information on the notebooks, and the entries and comments within each notebook. The primary function of *cynote* module is to allow for the creation of new notebooks and to add entries and comments into each notebook. Stored analysis results can also be displayed and stored as notebook entries. Each of these events will be logged in *db* database.

On the administrative end, this module allows for Table of Contents listing of each notebook, marking a notebook for archival and unarchival; thereby, controlling whether new entries can be added into a notebook. At the same time, it allows for users to notarize each entry and to generate digital signatures for each entry for detecting forced edits. Digital signature of each comment or entry is generated as a combination of MD5 and SHA hashes. Each of these events will be logged in *db* database.

**Module: *primer*.** This module implements computing features for the calculation of the annealing temperature of given DNA primers, taking account of monovalent ions (mainly sodium), divalent ions (mainly magnesium) and the concentration of the primers.

**Module: *savedatabase*.** This module allows users to backup the entire database and uploaded files to a '*cynote_database*' directory of an FTP server.

**Module: *sequence*.** This module imports BioPython library (Cock et al., 2009) to incorporate its available features to provide bioinformatics functions to perform restriction endonuclease digestion and BLAST-ing a given DNA sequence. A given DNA sequence can also be complemented, transcribed, translated, back transcribed, and back translated. A given amino acid sequence can be analyzed for amino acid composition and proportion, molecular weight, aromaticity, instability, flexibility, and isoelectric point.

**Module: *statistics*.** This module imports COPADS (Ling et al., 2009) to incorporate its available features to provide statistical analysis functions. Unlike primers and sequence analysis functions, results from statistical analysis will not be stored into *results* table for future retrieval.

### 3. Features of CyNote

Each of the 33 features currently in CyNote is categorized into one of the 4 functional categories (listed in Table 1), namely, general features, 21 CFR Part 11 features, bioinformatics features and statistical features. General and 21 CFR Part 11 features can be classified as note-taking features while bioinformatics and statistical features can be classified as computing features.

| **Feature Category: General functions** |
| --- |
| 1. Creating multiple notebooks to allow for project segregation |
| 2. Each (entry) notebook page can have multiple comments and/or file insertions |
| 3. Authorized users can authorize and de-authorize user(s) from the system |
| 4. Allow user to set a notebook to archive; thereby, locking it to prevent further insertion |
| 5. Method to streamline updating of CyNote and the underlying web2py |
| 6. Entire database (all the notebooks) can be backed-up onto a FTP server and date-time stamped |
| **Feature Category: 21 CFR Part 11** |
| 1. Users require to log-in before using CyNote |
| 2. Users are uniquely identified by user name and there can be no duplicate |
| 3. New user creation and user logins/logouts events are recorded |
| 4. Password for users is encrypted |
| 5. Each entry and comment is date-time stamped using UTC time |
| 6. The authorship of each entry and comment is logged |
| 7. User can notarize (affix a signature based on login) a specific entry |
| 8. Logging for authorizing/de-authorizing user(s) |
| 9. Logging for archiving/un-archiving notebooks |
| 10. Each entry into the database is numbered and any forced deletion will result in |
| subsequent errors |
| 11. There is no function to delete user, notebook, entry and comment |
| 12. Generate digital signature for each entry and comments for auditing and preventing forced editing |
| 13. Change of password function and logging of such events |
| 14. Force user to change password if last change is more than 30 days |
| **Feature Category: Bioinformatics** |
| 1. Transcribes from DNA to RNA |
| 2. Reverse-transcribes from RNA to DNA |
| 3. Translates from DNA/RNA to amino acid sequence |
| 4. Performs amino acid distribution, pI, MW, aromaticity, instability, flexibility, and secondary structure fractions on amino acid sequence |
| 5. Perform similarity search using NCBI BLAST |
| 6. Calculate annealing temperature of PCR primers |
| 7. Restriction enzyme mapping on a DNA sequence |
| **Feature Category: Statistics** |
| 1. Calculates descriptive statistics on a data sample |
| 2. Calculates linear regression line and tests Pearson's correlation from data with given correlation |
| 3. Calculates Phi correlation coefficient, Cohen's Kappa, P1-P2, Relative risk, and Odds ratio for 2x2 contingency table |
| 4. Z test for correlated proportions, Chi-square test with and without Yate's correction, McNemar's test with and without Yate's correction for 2x2 contingency table |
| 5. Calculates Goodman and Kruskal's gamma, Kendall's tau-a and Kendall's tau-c correlation for 2 x C contingency table |
| 6. Chi-square test for categorical data and R x C contingency table |

Table 1: Categorical Features of CyNote Version 1.3.

### 3.1. Use case: General note-taking

This use case demonstrates using CyNote for general note-taking and how auditing can be carried out.

After logging into CyNote, the user is able to create a new entry assuming that at least a notebook had been created (Figure 2). The newly created entry can be listed in "List all Entries" function which will list all entries in a reverse chronology order (Figure 3). A digital signature can be generated for the newly created entry, together with all entries in the entire system using the "Generate Digital Signature" function (Figure 4). This signature can be used to ensure that the entry is not changed by comparing subsequent signatures generated.

Figure 2: Creating a New Entry.

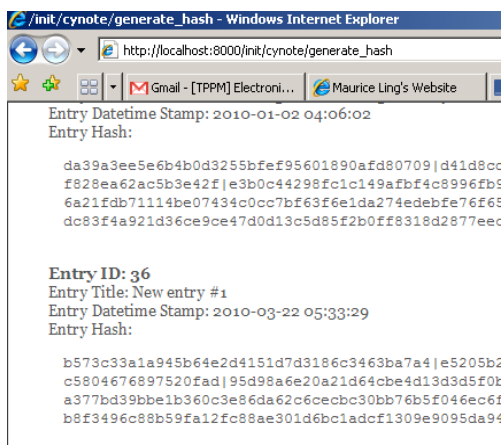

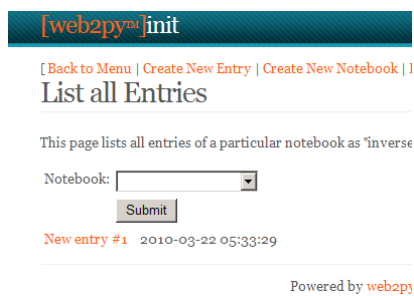

Figure 3: List all Entries.

Figure 4: Generated Digital Signatures.

## 3.2. Use case: Computing the annealing temperature of primers (Bioinformatics function)

This use case demonstrates the interplay of computing functions and the note-taking features. In this example, the annealing temperature for a pair of primers will be calculated and recorded as a new entry.

Analyzing the annealing temperature of primers is found on the Bioinformatics menu. The primer sequences are entered as shown in Figure 5. The results are shown (Figure 6) after clicking the "submit" button. The results are stored and can be displayed under the "Show all Results" function at the header bar (Figure 7). The results for entry into a notebook can be selected.

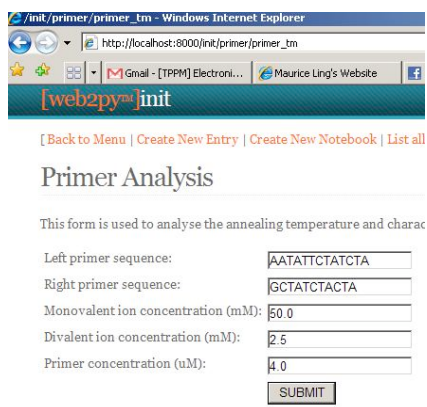

Figure 5: Analyzing Primers.

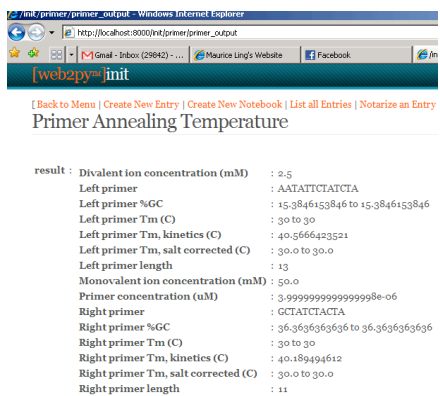

Figure 6: Primer Analysis Results.

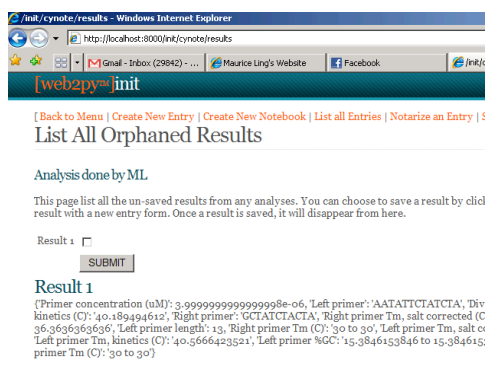

Figure 7: Show all Results.

A new entry form will be displayed for the user to copy and paste the results into the entry form (Figure 8). There is no automated entering of results into the notebook as a new entry as we believe in providing flexibility for users to record their results – as in a physical laboratory notebook.

Figure 8: Saving Results.

## 4. Evaluation against Controls for Electronic Records (Subpart B of 21 CFR 11)

There are a total of 12 controls for electronic records comprising of 11 controls for closed system and 1 additional control for open systems required under 21 CFR 11 (US FDA, 1997). CyNote may be considered as an open system as it require connection to publically available tools, such as NCBI BLAST, and the subsequent results are inserted into CyNote. This implies that part of CyNote database contents falls within the definition of open system which includes "an environment in which system access is not controlled by persons who are responsible for the content of electronic records that are on the system" (US FDA, 1997). We evaluate CyNote against each of the 12 control requirements as follows and summarized in Table 2 below:

(a) Validation of systems to ensure accuracy, reliability, consistent intended performance, and the ability to discern invalid or altered records.
   **CyNote:** Each database in CyNote uses an auto-increment number as primary key to number each record, such as entry, comments and event logs. At the same time, periodic digital signatures (a combination of SHA1, SHA225, SHA256, SHA384, SHA512 and MD5 hashes) of notebook entries and comments can be re-generated without deletion of the previous signatures of entries and comments. These imply that any deletion of record(s) will be identified by missing incremental record number and alteration of record(s) can be identified by comparing historical digital signatures of the suspecting record.

(b) The ability to generate accurate and complete copies of records in both human readable and electronic form suitable for inspection, review, and copying by the agency.
   **CyNote:** Complete databases can be exported into human readable text file by a function provided by web2py.

(c) Protection of records to enable their accurate and ready retrieval throughout the records retention period.
   **CyNote:** Multiple historical digital signatures, depending on institutional policy, can be generated for each notebook record to ensure record accuracy. These signatures can be exported as into human readable text file by a function provided by web2py.

(d) Limiting system access to authorized individuals.
   **CyNote:** CyNote implements access control which requires users to log in before use.

(e) Use of secure, computer-generated, time-stamped audit trails to independently record the date and time of operator entries and actions that create, modify, or delete electronic records. Record changes shall not obscure previously recorded information. Such audit trail documentation shall be retained for a period at least as long as that required for the subject electronic records and shall be available for agency review and copying.
   **CyNote:** Each notebook entry is date and time-stamped using Python's UTC time module and there are no functions for record amendments or deletion.

(f) Use of operational system checks to enforce permitted sequencing of steps and events, as appropriate.
   **CyNote:** There are 3 main operational checks built into CyNote on top of event logging. Firstly, user must log in and the log in detailed will be checked for existence and correctness in the users' database, and the user must be authorized by another authorized user to use the system. Secondly, a notebook must be created before entries into the specific notebook can take place. Lastly, a notebook entry must be present before comments can be linked to the specific entry.

(g) Use of authority checks to ensure that only authorized individuals can use the system, electronically sign a record, access the operation or computer system input or output device, alter a record, or perform the operation at hand.
   **CyNote:** Every entry into the database will be tagged to a user name which can be cross-referenced with login/logout table.

(h) Use of device (e.g., terminal) checks to determine, as appropriate, the validity of the source of data input or operational instruction.
   **CyNote:** Appropriate data format checking is done in the data-entry forms, such as notebook entry/comments entry.

(i) Determination that persons who develop, maintain, or use electronic record/electronic signature systems have the education, training, and experience to perform their assigned tasks.
   **CyNote:** This requirement depends on the training and educational policy of the user and cannot be fulfilled independently by CyNote.

(j) The establishment of, and adherence to, written policies that hold individuals accountable and responsible for actions initiated under their electronic signatures, in order to deter record and signature falsification.
   **CyNote:** This requirement depends on institutional policy within which CyNote operates and cannot be fulfilled independently by CyNote.

(k) Use of appropriate controls over systems documentation including: (1) Adequate controls over the distribution of, access to, and use of documentation for system operation and maintenance. (2) Revision and change control procedures to maintain an audit trail that documents time-sequenced development and modification of systems documentation.
   **CyNote:** This requirement depends on institutional policy within which CyNote operates and cannot be fulfilled independently by CyNote.

(l) Persons who use open systems to create, modify, maintain, or transmit electronic records shall employ procedures and controls designed to ensure the authenticity, integrity, and, as appropriate, the confidentiality of electronic records from the point of their creation to the point of their receipt.
   **CyNote:** The open aspects of CyNote as a result of linking to external databases does not read data from or write into the notebook tables. In addition, based on web2py documentation (Di Pierro, 2009), security measures had been implemented to protect confidentiality of the system. For further security and confidentiality, CyNote should be used as a personal laboratory notebook replacement (operating off a personal computer) rather than deploying for team use.

| 21 CFR Part 11 Subpart B Requirements | CyNote Compliance Mechanism |
|---|---|
| (a) Validation of systems to ensure accuracy, reliability, consistent intended performance, and the ability to discern invalid or altered records. | Event logging and digital signatures of records. |
| (b) The ability to generate accurate and complete copies of records in both human readable and electronic form suitable for inspection, review, and copying by the agency. | Export records as text file. |
| (c) Protection of records to enable their accurate and ready retrieval throughout the records retention period. | Multiple historical digital signatures. |
| (d) Limiting system access to authorized individuals. | Log in procedure. |
| (e) Use of secure, computer-generated, time-stamped audit trails to independently record the date and time of operator entries and actions that create, modify, or delete electronic records. | Date time stamp. |
| (f) Use of operational system checks to enforce permitted sequencing of steps and events, as appropriate. | Log in procedure, every comment must be tagged to an entry and every entry must be tagged to a notebook. |
| (g) Use of authority checks to ensure that only authorized individuals can use the system, electronically sign a record, access the operation or computer system input or output device, alter a record, or perform the operation at hand. | Tagging every entry to a specific logged in user. |
| (h) Use of device (e.g., terminal) checks to determine, as appropriate, the validity of the source of data input or operational instruction. | Data checking in entry forms. |
| (i) Determination that persons who develop, maintain, or use electronic record/electronic signature systems have the education, training, and experience to perform their assigned tasks. | Dependent on organizational policy – cannot be fulfilled by CyNote. |
| (j) The establishment of, and adherence to, written policies that hold individuals accountable and responsible for actions initiated under their electronic signatures, in order to deter record and signature falsification. | Dependent on organizational policy – cannot be fulfilled by CyNote. |
| (k) Use of appropriate controls over systems documentation including: (1) Adequate controls over the distribution of, access to, and use of documentation for system operation and maintenance. (2) Revision and change control procedures to maintain an audit trail that documents time-sequenced development and modification of systems documentation. | Dependent on organizational policy – cannot be fulfilled by CyNote. |
| (l) Persons who use open systems to create, modify, maintain, or transmit electronic records shall employ procedures and controls designed to ensure the authenticity, integrity, and, as appropriate, the confidentiality of electronic records from the point of their creation to the point of their receipt. | Using CyNote as a standalone system. |

Table 2: <u>Summary of 21 CFR Part 11 Subpart B compliance by CyNote.</u>

**5. Evaluation against Controls for Identification Codes/Passwords (Subpart C of 21 CFR 11)**

There are a total of 5 controls for identification codes and passwords required under 21 CFR 11 (US FDA, 1997). We evaluate CyNote against each of the 5 control requirements as follows and summarized in Table 3 below:

(a) Maintaining the uniqueness of each combined identification code and password, such that no two individuals have the same combination of identification code and password.
> **CyNote:** The user database ensures that each user name (identification code) is unique; hence, no two individuals have the same combination of identification code and password.

(b) Ensuring that identification code and password issuances are periodically checked, recalled, or revised (e.g., to cover such events as password aging).
> **CyNote:** The user is forced to change password before continuation if last change is more than 90 days.

(c) Following loss management procedures to electronically deauthorize lost, stolen, missing, or otherwise potentially compromised tokens, cards, and other devices that bear or generate identification code or password information, and to issue temporary or permanent replacements using suitable, rigorous controls.
> **CyNote:** This requirement depends on institutional policy within which CyNote operates and cannot be fulfilled independently by CyNote. However, CyNote allows users to be deauthorized.

(d) Use of transaction safeguards to prevent unauthorized use of passwords and/or identification codes, and to detect and report in an immediate and urgent manner any attempts at their unauthorized use to the system security unit, and, as appropriate, to organizational management.

**CyNote:** Every erroneous logins will be logged with the number of erroneous attempts.

(e) Initial and periodic testing of devices, such as tokens or cards, that bear or generate identification code or password information to ensure that they function properly and have not been altered in an unauthorized manner.

**CyNote:** All passwords changes, whether successful or not, will be logged as events. When changes are unsuccessful, the reasons will be logged.

| 21 CFR Part 11 Subpart C Requirements | CyNote Compliance Mechanism |
|---|---|
| (a) Maintaining the uniqueness of each combined identification code and password, such that no two individuals have the same combination of identification code and password. | Every user must have unique user name. |
| (b) Ensuring that identification code and password issuances are periodically checked, recalled, or revised (e.g., to cover such events as password aging). | Force user to change password if more than 90 days has lapsed since last change. |
| (c) Following loss management procedures to electronically deauthorize lost, stolen, missing, or otherwise potentially compromised tokens, cards, and other devices that bear or generate identification code or password information, and to issue temporary or permanent replacements using suitable, rigorous controls. | Dependent on organizational policy – cannot be fulfilled by CyNote. |
| (d) Use of transaction safeguards to prevent unauthorized use of passwords and/or identification codes, and to detect and report in an immediate and urgent manner any attempts at their unauthorized use to the system security unit, and, as appropriate, to organizational management. | Errors in log in will be tracked. |
| (e) Initial and periodic testing of devices, such as tokens or cards, that bear or generate identification code or password information to ensure that they function properly and have not been altered in an unauthorized manner. | Password change events will be logged. |

Table 3: <u>Summary of 21 CFR Part 11 Subpart C compliance by CyNote.</u>

## 6. Discussion

As an academic once said that "*science is about paying attention to details*" and I believe that a significant part of it lies in good documentation as reproducibility is paramount in any scientific work. Traditionally, research documentation is carried out using bounded paper journals and pen where audit trail can be visually and logically established. However, advances in technology had generated a large volume of electronic records rendering traditional documentation methods ineffective. Hence, a class of systems known as Electronic Laboratory Notebook (ELN) is created to address this concern (Butler, 2005, Tabard et al., 2008). CyNote is an ELN aiming to provide electronic records with the same level of audit trail as traditional documentation by adhering to US FDA 21 CFP Part 11 (US FDA, 1997).

The ideological foundation of CyNote is a blog where comments can be appended onto each entry (Brady, 2005). The advantage of this model is that each experiment can be written up as an entry and any follow ups can be appended as comments. Unlike traditional paper-based documentation where navigating follow up experiments can span across non-consecutive pages, navigating an experiment on CyNote can be as easy as reading a blog entry with comments.

The most similar ELN to CyNote is ORNL Electronic Notebook (http://www.csm.ornl.gov/~geist/java/applets/enote/) which also open-sourced and followed a blog structure. Although ORNL's system maintained a list of authorized users, it allows for users to edit entries. In addition, ORNL's system does not implement logging of events and there is no digital signature generated. The only audit clues provided is the date time stamp on individual files which can be easily manipulated. This suggests that it may be possible to amend or even delete an entry in ORNL's system without detection. Hence, ORNL Electronic Notebook is unlikely to conform to the requirement Subpart B(a) of FDA 21 CFP Part 11 (US FDA, 1997). Current work in CyNote focused largely on the compliance of 21 CFP Part 11 (US FDA, 1997). We had evaluated CyNote on each of the guidelines in Part B and C of 21 CFP Part 11, and found our system to be generally compliant. However, certain rules such as Part B Rule J which states that "*The establishment of, and adherence to, written policies that hold individuals accountable and responsible for actions initiated under their electronic signatures, in order to deter record and signature falsification*", defines institutional policy and cannot be fully complied by any computerized system. This suggests that CyNote or any ELN systems can only fulfil part of the compliance.

CyNote was developed as pilot software for students' use to manage their projects. As projects can vary greatly, the needs may differ greatly. Hence, there must be a realistic scope for implementation and room for future development (Prism Forum, 2004). Current bioinformatics and statistical modules were developed using the model-view-controller (MVC) paradigm. We believe that CyNote can be extended by other modules can be built using the same paradigm in future.

## 7. References

Brady, M. 2005. Blogging, personal participation in public knowledge-building on the web. Chimera Working Paper 2005-02. Colchester: University of Essex.

Butler, D. 2005 Electronic Notebooks: A new leaf. Nature 436:20-21.

Cock, P. J., Antao, T., Chang, J. T., Chapman, B. A., Cox, C. J., Dalke, A., Friedberg, I., Hamelryck, T., Kauff, F., Wilczynski, B. & de Hoon, M. J. (2009) Biopython: freely available Python tools for computational molecular biology and bioinformatics. Bioinformatics, 25, 1422-3.

Di Pierro, M. 2009. Web2py: enterprise web framework, 2$^{nd}$ edition. Wiley Publishers.

Frey, J. 2008. Curation of laboratory experimental data as part of the overall data lifecycle. International Journal of Digital Curation 3(1): 44-62.

Garabedian, TE. 1997. Laboratory record keeping. Nature Biotechnology 15: 799-800.

Garabedian, TE. 2003. Recent developments in intellectual property law: avoiding traps in the pursuit of university reseach. Reseach Management Review, Sping/Winter 2003.

Ling, MHT. 2009. COPADS: Collection of Python Algorithms and Data Structures. http://copads.sourceforge.net

Kanare, HM. 1985. Writing the laboratory notebook. American Chemical Society.

Newman, C. 2004. SQLite (Developers' Library). Sams Publishing.

Prism Forum. 2004. Electronic Laboratory Notebooks. A Prism Forum White Paper. http://www.prismforum.org/images/ELN%20WHITE%20PAPER%20v2.1.pdf [Last accessed: 1st February 2010]

Schreier, AA., Wilson, K, Resnik, D. 2006. Academic Research Record-Keeping: Best Practices for Individuals, Group Leaders, and Institutions. Academic Medicine 81(1): 42-47.

Tabard, A., Mackay, WE., Eastmond, E. 2008. From individual to collaborative: the evolution of prism, a hybrid laboratory notebook. Proceedings of the ACM 2008 conference on Computer supported cooperative work.

US FDA. 1997. Electronic records; Electronic signatures (21 CFR Part 11). Federal Register 62(54):13430-13466. [Docket No. 92N–0251]

US NIH. 2008. Guidelines for SCIENTIFIC RECORD KEEPING in the Intramural Research Program at the NIH, 1st edition. http://www1.od.nih.gov/oir/sourcebook/ethic-conduct/RECORDKEEPING.pdf. [Accessed on November 11, 2009]

William, M., Bozyczko-Coyne, D., Dorsey, B, Larson, S. 2008. Laboratory notebooks and data storage. Current Protocols Essential Laboratory Techniques, Appendix 2A.